\documentclass[twocolumn]{aastex63}
\usepackage{lineno}
\usepackage{epstopdf}
\shorttitle{Unstable Gas Supply in CL-AGNs}
\shortauthors{Wang et al.}

\begin{document}

\title{Instability of Circumnuclear Gas Supply as An Origin of ``Changing-look'' Phenomenon of Supermassive Blackholes}

\correspondingauthor{J. Wang \& D. W. Xu}
\email{wj@nao.cas.cn, dwxu@nao.cas.cn}

\author{J. Wang}
\affiliation{Guangxi Key Laboratory for Relativistic Astrophysics, School of Physical Science and Technology, Guangxi University, Nanning 530004,
People's Republic of China}
\affiliation{Key Laboratory of Space Astronomy and Technology, National Astronomical Observatories, Chinese Academy of Sciences, Beijing 100101,
People's Republic of China}
\affiliation{GXU-NAOC Center for Astrophysics and Space Sciences, Nanning, 530004, People's Republic of China}

\author{D. W. Xu}
\affiliation{Key Laboratory of Space Astronomy and Technology, National Astronomical Observatories, Chinese Academy of Sciences, Beijing 100101,
People's Republic of China}
\affiliation{School of Astronomy and Space Science, University of Chinese Academy of Sciences, Beijing, People's Republic of China}

\author{Xinwu Cao}
\affiliation{Institute for Astronomy, School of Physics, Zhejiang University, 866 Yuhangtang Road, Hangzhou 310058, People's Republic of China}

\author{C. Gao}
\affiliation{Guangxi Key Laboratory for Relativistic Astrophysics, School of Physical Science and Technology, Guangxi University, Nanning 530004,
People's Republic of China}
\affiliation{Key Laboratory of Space Astronomy and Technology, National Astronomical Observatories, Chinese Academy of Sciences, Beijing 100101,
People's Republic of China}
\affiliation{School of Astronomy and Space Science, University of Chinese Academy of Sciences, Beijing, People's Republic of China}

\author{C. H. Xie}
\affiliation{Guangxi Key Laboratory for Relativistic Astrophysics, School of Physical Science and Technology, Guangxi University, Nanning 530004,
People's Republic of China}
\affiliation{Key Laboratory of Space Astronomy and Technology, National Astronomical Observatories, Chinese Academy of Sciences, Beijing 100101,
People's Republic of China}
\affiliation{School of Astronomy and Space Science, University of Chinese Academy of Sciences, Beijing, People's Republic of China}

\author{J. Y. Wei}
\affiliation{Key Laboratory of Space Astronomy and Technology, National Astronomical Observatories, Chinese Academy of Sciences, Beijing 100101,
People's Republic of China}
\affiliation{School of Astronomy and Space Science, University of Chinese Academy of Sciences, Beijing, People's Republic of China}






\begin{abstract}

The origin of the ``Changing-look'' (CL) phenomenon in supermassive black holes (SMBHs) remains an open issue. This study aims to shed light on this phenomenon 
by focusing on a sample that encompasses 
all known repeating CL active galactic nuclei (AGNs). 
Through the identification of a characteristic time scale for the CL phenomenon, it was observed that larger SMBHs possess shorter characteristic timescales, while smaller SMBHs exhibit longer timescales. These findings reveal a significant contrast 
to the traditional AGN variability that has been adequately explained by the AGN's disk instability model. This stark discrepancy highlights a distinct origin of the CL phenomenon, distinguishing it from traditional AGN variability.
By properly predicting the characteristic time scale and its dependence on SMBH mass, we propose that the CL phenomenon is likely a result of a variation in accretion rate caused by a sudden change in the supply of circumnuclear gas during the transition between active and passive SMBH fueling stages.
\end{abstract}



\section{Introduction} \label{sec:intro}

``Changing-look'' active galactic nuclei (CL-AGNs), a rare phenomenon revealed by repeat spectroscopic observations, are characterized by a temporary transition in their spectral types. This transition pertains to the appearance or disappearance of their broad emission lines, ranging from type I, intermediate type, and type II over a time scale spanning several years or even decades (Ricci \& Trakhtenbrot 2022; Komossa \& Grupe 2023). The identification of CL-AGNs has surged dramatically due to recent advancement in multi-epoch spectroscopy. However, the number of confirmed CL-AGNs remains limited, with $\sim$400 instances identified (e.g., L{\'o}pez-Navas et al. 2022, 2023; Guo et al. 2023; Wang et al. 2023a; Wang et al. 2024; Zeltyn et al. 2024).
Out of these, only ten repeating CL-AGNs (RCL-AGNs) have exhibited repeated type transitions (Wang et al. 2020, 2022, Marin et al. 2019 and references therein). 

Although there is competing evidence supporting the CL phenomenon is likely
due to a variation in accretion rate of a supermassive blackhole (SMBH, e.g., Sheng et al. 2017, 2020; MacLeod et al. 2019; Yang et al. 2018; 
Feng et al. 2021)\footnote{Other possible interpretations include: for instance, an accelerating outflow (e.g., Shapovalova et al. 2010), 
a variation of the obscuration (e.g., Elitzur 2012), and a tidal disruption event (e.g., Merloni et al. 2015; Blanchard et al.
2017; Padmanabhan \& Loeb 2021; Wang et al. 2023b). }, 
the physics behind this change in accretion rate remains an open question
(e.g., Nagoshi \& Iwamuro 2022; Panda \& Sniegowska 2022). 
There is a viscosity crisis in understanding of the phenomenon in the context of the standard
Shakura-Sunyaev disk (SSD) model (Shakura \& Sunyaev 1973) in which the expected viscous timescale of optical emission coming from the outer accretion disk
is larger than the timescale of the observed CL phenomenon by an order of
magnitude (Lawrence 2018 and references therein).  Based upon a revision of the SSD model,
this crisis could be theoretically alleviated by introducing a local disk thermal 
instability (e.g., Husemann et al. 2016) or a magnetic field (e.g., Ross et al. 2018; 
Stern et al. 2018; Dexter \& Begelamn 2019; Pan et al. 2021; Feng et al. 2021; Cao et al. 2023).

The verification of these models through observations poses a significant challenge, as the majority of CL-AGNs are typically observed as isolated events, with the exception of the rare occurrence of ten RCL-AGNs that exhibit at least one complete type transition cycle. Studying these specific RCL-AGNs is therefore of particular importance as they provide valuable opportunities for identifying the characteristic timescale linked to the CL phenomenon. In this study, we focus on a sample that comprises all of the well-known RCL-AGNs, aiming to unveil the origin of the CL phenomenon and examine whether the CL phenomenon differs in origin from the traditional variability observed in AGNs, which has already been suitably explained by the disk instability (DI) model (e.g., Kawaguchi et al. 1998; Ulrich et al. 1997).

The paper is organized as follows. Sections 2 and 3 present the sample of RCL-AGNs and statistical results, respectively.
The implications are described in Section 4. 
A $\Lambda$CDM cosmological model with parameters H$_0=70\,\mathrm{km\,s^{-1}\,Mpc^{-1}}$, $\Omega_{\mathrm{m}}=0.3$, and
$\Omega_\Lambda=0.7$ is adopted throughout.

\section{RCL-AGNs Sample} \label{sec:style}

To date, only ten RCL-AGNs have been well identified.
They are  Mrk\,590, Mrk\,1018, 
NGC\,1566, 
NGC\,4151, NGC\,7603, Fairall\,9, 3C\,390.3, UGC\,3223 and B3\,0749+460A 
(e.g., Wang et al. 2020, 2022; Marin et al. 2019 and references therein).

Table 1 lists the basic properties of the 11 RCL-AGNs, including the viral SMBH mass ($M_{\mathrm{BH}}$)
and Eddington ratio $L_{\mathrm{bol}}/L_{\mathrm{Edd}}$, where $L_{\mathrm{bol}}$ and 
$L_{\mathrm{Edd}}=1.5\times10^{38}(M_{\mathrm{BH}}/M_\odot)$ are the bolometric and Eddington 
luminosity (Netzer 2013), respectively. 
The type transition sequence is shown in Figure 1 for the 11 RCL-AGNs, 
after combining the results compiled 
from literature 
\footnote{For NGC~1566, a few spectra taken in 
early 1970's in Alloin et al. (1986) are excluded because of their quite low signal-to-noise 
ratio ($\mathrm{S/N}<15$).}
and our new 
spectroscopy of five out of the 11 RCL-AGNs presented in this paper 
(See Appendix A for the details). 

Based on the H$\alpha$ and H$\beta$ emission line profiles, 
the ``turn-on'' state denotes an object has a spectrum with an evident 
H$\beta$ broad emission, i.e., early than the Seyfert-1.5 type. Instead, a spectrum with 
weak or absent H$\beta$ broad emission, i.e., late than the Seyfert-1.5 type, is referred as 
the ``turn-off'' state. For B3\, 0749+460A, the classical Balmer broad emission line, rather than 
the double-peaked emission, is adopted to distinguish the ``turn-on'' and ``turn-off'' states 
(Wang et al. 2022).

\begin{table*}
        \centering
        \caption{Properties of the ten RCL-AGNs}
        \label{tab:example_table}
        \begin{tabular}{ccccccc} 
        \hline
        \hline
        Object & $z$ &  $\log(M_{\mathrm{BH}}/M_\odot)$  & $L/L_{\mathrm{Edd}}$ & $\Delta t_{\mathrm{min}}$ & $\Delta t_{\mathrm{max}}$ & References\\
              &     &     &     & yr  & yr \\
          (1) & (2) & (3) & (4) & (5) & (6) & (7)\\     
        \hline     
       NGC\,1566 & 0.00502 & 6.9 & 0.05 & 5 & 19 & 1,2,3,4,23\\
       NGC\,4151 & 0.00333 & 7.4 & 0.04 & 10 & 17 & 5,6,7,24,25,27\\
       Mrk\,590 & 0.02638 & 7.6 & 0.05 &  8 & 27 & 8.9,25\\
       Mrk\,1018 & 0.04296 & 7.7 & 0.06 & 25 & 36 & 10,11,25,26\\
       UGC\,3223 & 0.01562 & 7.7 & 0.014 & 9 & 32 & 12,13,25\\
       B3\,0749+460A & 0.05174 & 8.1 & 0.012 & 3 & 17 & 14,15\\
       SDSS\,J162829.17+432948.5 & 0.2603 & 8.2 & 0.03 & 1 & 6 & 16,31\\ 
       Fairall\,9 & 0.04614 & 8.3 & 0.013 & 3 & 6 & 17,18,28\\
       SDSS\,J132457.29+480241.3 &0.271847 & 8.4 & 0.06 & 1 & 9 & 29,30,31,32,33\\  
       NGC\,7603 & 0.02876 & 8.4 & 0.03 & 1 & 2 & 19,20,25\\
       3C\,390.3 & 0.05613 & 9.3 & 0.02 & 4 & 10 & 5,21,22,25\\
        \hline
        \end{tabular}
        \tablecomments{Columns (1): Object IAU name; Column (2): Redshift; 
        Columns (3) and (4): SMBH viral mass and Eddington ratio 
        $L_{\mathrm{bol}}/L_{\mathrm{Edd}}$,
        where $L_{\mathrm{bol}}$ and $L_{\mathrm{Edd}}$ are bolometric and Eddington luminosity
        ($L_{\mathrm{Edd}}=1.5\times10^{38}(M_{\mathrm{BH}}/M_\odot)$), 
        respectively; Columns (5) and (6): the maximum and minimum of CL cycle time in unit of year, see the main text for the definitions;
        Column (7): 1: Pastoriza \& Gerola (1970); 2: Alloin et al. (1985); 3: Alloin et al. (1986);
        4: Oknyansky et al. (2019); 5: Penston \& Perez (1984); 6: Malkov et al. (1997); 7: Shapovalova et al. (2008); 8: Denney et al. (2014); 9: Raimundo et al. (2019); 10: Cohen et al. (1986); 11: McElroy et al. (2016); 12: Stirpe (1990); 13: Wang et al. (2020); 14: Hon et al. (2020); 15: Wang et al. (2022); 16: Zeltyn et al. (2022); 17: Kollatschny \& Fricke (1985); 18: Stirpe et al. (1989); 19: Tohline \& Osterbrock (1976); 20: Kollatschny et al. (2000); 21: Veilleux \& Zheng (1991); 22: Sergeev et al. (2017); 23: Oknyansky et al. (2020); 24: Chen et al. (2023); 25: This study; 26: Osterbrock (1981), 27: Osterbrock (1977); 28: Han et al. (2011); 29: Zeltyn et al. (2024); 30: MacLeod et al. (2016); 31: MacLeod et al. (2019); 32: Hutsem{\'e}kers et al. (2019); 33: Ruan et al. (2016)}
\end{table*}

\begin{figure*}[ht!]
\plotone{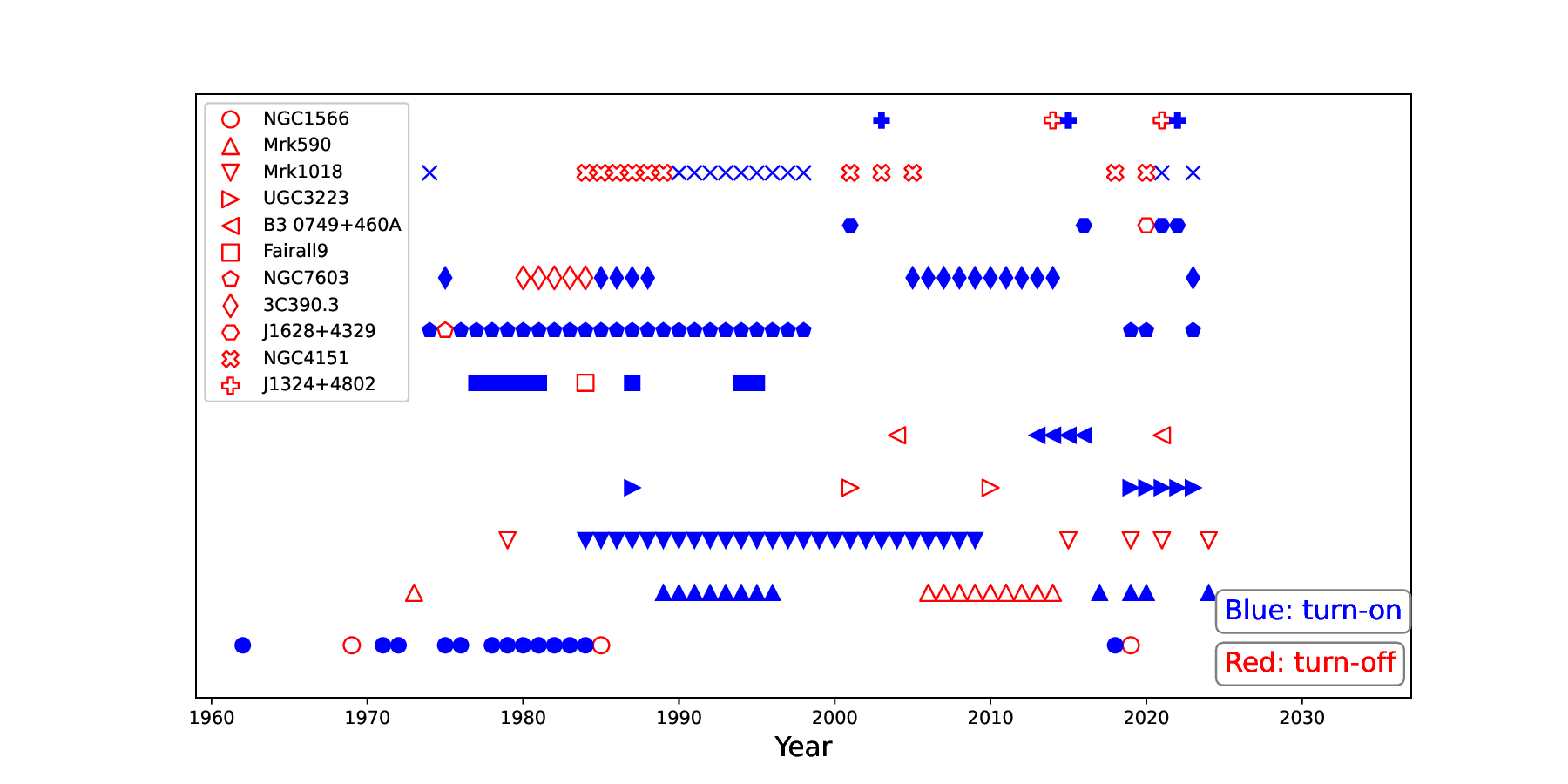}
\caption{Spectral type sequence of the 11 RCL-AGNs. 
Different objects are shown by the 
different symbols, and the ``turn-on'' and ``turn-off'' states are denoted by the 
blue-solid and red-open symbols, respectively.
\label{fig:general}}
\end{figure*}

\section{A Dependence of CL Phenomenon on SMBH Mass}

\subsection{CL Cycle Time}
One can see from Figure 1 that the spectral type sequences of the 11 RCL-AGNs are 
extremely under sampled, 
which leads to a difficulty in definition of the characteristic time scale of CL phenomenon. 
A CL cycle time $\Delta t_{\mathrm{cyc}}$ is adopted here to address this difficulty.
$\Delta t_{\mathrm{cyc}}$ is defined as 
a duration spent in a spectral type sequence of $A\rightarrow B\rightarrow A$, where $A$ and $B$ denote
either ``turn-on'' or ``turn-off'' state and $A\neq B$. The exact value of 
$\Delta t_{\mathrm{cyc}}$ is still hard to be determined because of the extremely under sampling of 
the transition sequence. 

Let $A\rightarrow B\rightarrow A=(...,A_n)\rightarrow (B_1,...B_m)\rightarrow (A_{n+1},...)$, where $X_i$ ($X=A$ or $B$) denotes the $i$th spectrum with a spectral type of $X$ obtained at time $t_{X_i}$.  
We then define the maximum of $\Delta t_{\mathrm{cyc}}$ as
\begin{equation}
  \Delta t_{\mathrm{max}}=t_{A_{n+1}}-t_{A_n}
\end{equation}

The corresponding minimum is defined as 
\begin{equation}
  \Delta t_{\mathrm{min}}=
   \left\{\begin{array}{ll}
         t_{B_m}-t_{B_1}, & \mathrm{if}\ m\geq2; \\ 
         1\ \mathrm{yr},  & \mathrm{if}\ m=1;
        \end{array}
   \right.
\end{equation}
These definitions are illustrated in the lower panel of Figure 2. 
For an object with multi CL cycles (i.e., multiple $\Delta t_{\mathrm{max}}$ and $\Delta t_{\mathrm{min}}$), the averaged $\Delta t_{\mathrm{max}}$ and 
$\Delta t_{\mathrm{min}}$ are finally used in the subsequent statistical study.
The determined $\Delta t_{\mathrm{max}}$ and $\Delta t_{\mathrm{min}}$ are listed in Table 1 for 
each of 11 RCL-AGNs.

\begin{figure*}
\plotone{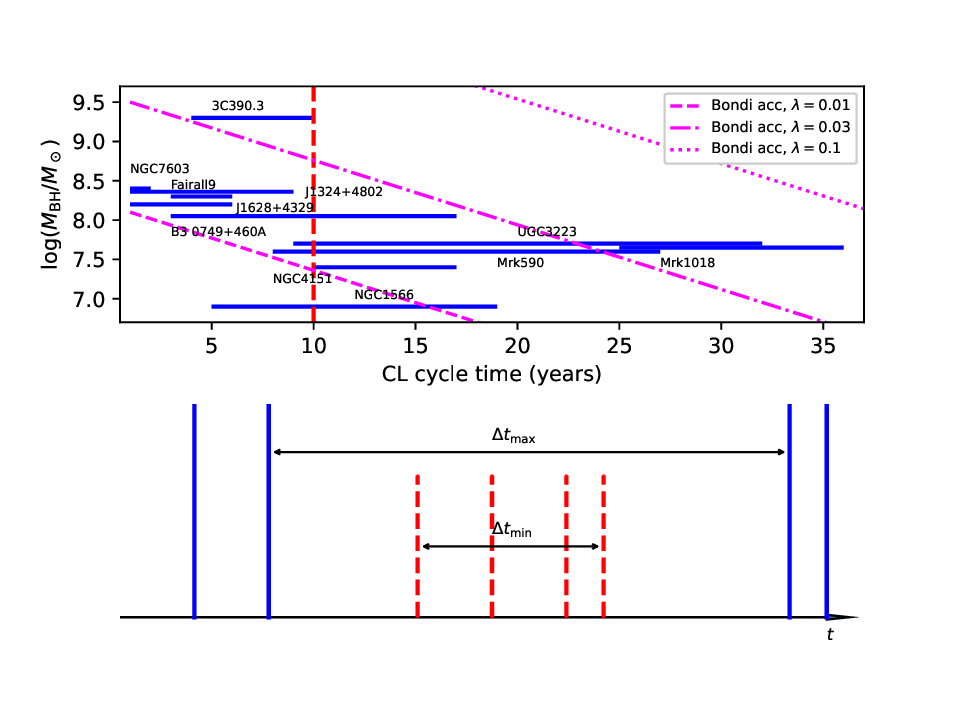}
\caption{\it Lower panel: \rm An illustration of definitions of the upper and lower limits of
CL cycle time. i.e., $\Delta t_{\mathrm{max}}$ and $\Delta t_{\mathrm{min}}$. The horizontal axis represents the direction of time. The vertical lines mark the time 
when a spectrum is obtained. The same spectral type is shown by the same line style and color.
\it Upper panel: \rm A dependence of $\Delta t_{\mathrm{cyc}}$ on $M_{\mathrm{BH}}$ revealed in 
the 11 RCL-AGNs. The coverage between $\Delta t_{\mathrm{min}}$ and $\Delta t_{\mathrm{max}}$ is 
shown by a horizontal bar for each of the RCL-AGNs. 
The vertical red-dashed lines marks $\Delta t_{\mathrm{max}}=10$yr for 3C\,390.3.
The magenta lines show 
the theoretical $M_{\mathrm{BH}}-\Delta t$ relations with different 
$\lambda=L_{\mathrm{bol}}/L_{\mathrm{Edd}}$ (see Eq. (3) and Eq. (D16)), which are 
predicted from Bondi accretion model with a sound speed of $10\ \mathrm{km\ s^{-1}}$ 
and surrounding gas with a radius $\propto L^{1/2}$. The dot-dashed line denotes the case with   
$\lambda=0.03$, the average value of the current RCL-AGN sample. 
\label{fig:general}}
\end{figure*}

\subsection{$\Delta t_{\mathrm{cyc}}$ versus $M_{\mathrm{BH}}$}

Based on the determined $\Delta t_{\mathrm{max}}$ and $\Delta t_{\mathrm{min}}$, 
the upper panel of Figure 2 presents a dependence of $\Delta t_{\mathrm{cyc}}$ on 
SMBH viral mass ($M_{\mathrm{BH}}$)
for the 11 RCL-AGNs. Each of the RCL-AGNs is shown by a horizontal bar covering a 
range between $\Delta t_{\mathrm{min}}$ and $\Delta t_{\mathrm{max}}$. 
One should be bear in mind that most of these ranges are far larger 
than the real ranges of $\Delta t_{\mathrm{cyc}}$ 
according to the definitions and the under sampled spectral type sequences. 

An inspection of the plot enables us to identify a trend in which 
the three RCL-AGNs with a massive SMBH 
(i.e., $M_\mathrm{BH}>10^8M_\odot$) tends to have a short CL cycle time 
(i.e., $\Delta t_{\mathrm{max}}<10$\ yr). On the contrary, the remain RCL-AGNs with
a less massive SMBH
show relatively large $\Delta t_{\mathrm{cyc}}$. The values of $\Delta t_{\mathrm{max}}$ of 
the three RCL-AGNs with a massive SMBH are, in fact, 
either smaller than or comparable to the values of 
$\Delta t_{\mathrm{min}}\sim7-10$\ yr
of the other RCL-AGNs with $M_{\mathrm{BH}}<10^8M_\odot$.

In addition, as shown in the top panel of Figure 3, there is a similar, but degraded 
dependence of $\Delta t_{\mathrm{cyc}}$ on $L_{\mathrm{bol}}$,  
which is not surprising due to their comparable 
$L_{\mathrm{bol}}/L_{\mathrm{Edd}}$ as shown in Table 1.

\section{Implications}

\subsection{Disk Instability}

Although the source is still under debate, it is widely accepted that
the traditional AGN variability is resulted 
from the instability occurring in the accretion disk surrounding the central SMBH.
Based on the quantitative statistical predictions for the spectrum of fluctuation for 
different models, comparisons with observations favor the DI model over 
the starburst model (e.g., Kawaguchi et al. 1998; Hawkins 2002).

Nevertheless, the $M_{\mathrm{BH}}-\Delta t_{\mathrm{cyc}}$ dependence revealed in a sample of 
RCL-AGNs is hard to be understood in the DI model on both 
theoretical and observational grounds.


On the observational ground, it is long known that more massive SMBHs have larger 
characteristic variability time scales than do less massive SMBHs. 
The Seyfert galaxies are well known to vary in brightness on a time scale of a month or less,
and radio-quiet quasars on a time scale of months or more 
(e.g., Hawkins et al. 1996;
Cristiani et al. 1996; Hawkins 2002; Sun et al. 2018; Zachary et al. 2022; De Cicco et al. 2022; 
Graham et al. 2020; see Ulrich et al. 1997 for a review).

On the theoretical ground, in the context of the SSD model, 
possible scenarios adopted to understand CL phenomenon include 
classical viscous radial inflow (e.g., 
Gezari et al. 2017) or thermal instability (e.g., Siemiginowska et al. 1996). 
The time scales predicted in both scenarios, however, conflict with the $M_{\mathrm{BH}}-\Delta t_{\mathrm{cyc}}$ dependence shown in the upper panel of Figure 2. Specifically speaking, 
the characteristic time scales are inferred to be 
$t_{\mathrm{infl}}\propto L^{-5/6}M_{\mathrm{BH}}^{2/3}=
L^{-1/6}(L/L_{\mathrm{Edd}})^{-2/3}$ and $t_{\mathrm{th}}\propto L^{1/2}$ for 
the viscous and thermal instabilities, respectively.
when a variation is observed at a given wavelength (See Appendix B for the details).
Taking into account of the comparable $L/L_{\mathrm{Edd}}$ of the 11 RCL-AGNs, 
both characteristic times scales are implied to be independent on $M_{\mathrm{BH}}$.  
In addition, there is no clear relation between both characteristic times scales and 
the estimated $\Delta t_{\mathrm{cyc}}$ as shown in the top and middle panels of Figure 3.

Finally, an undeniable fact is that
CL phenomenon could be instead resulted from a variation of obscuration. The crossing 
time can be estimated as $t_{\mathrm{cross}}\propto L^{3/4}M_{\mathrm{BH}}^{-1/2}$ (see
Appendix C for the details), which is found to be
not related with $\Delta t_{\mathrm{cyc}}$ again as shown by the bottom panel of Figure 3.

\begin{figure}[htp!]
\plotone{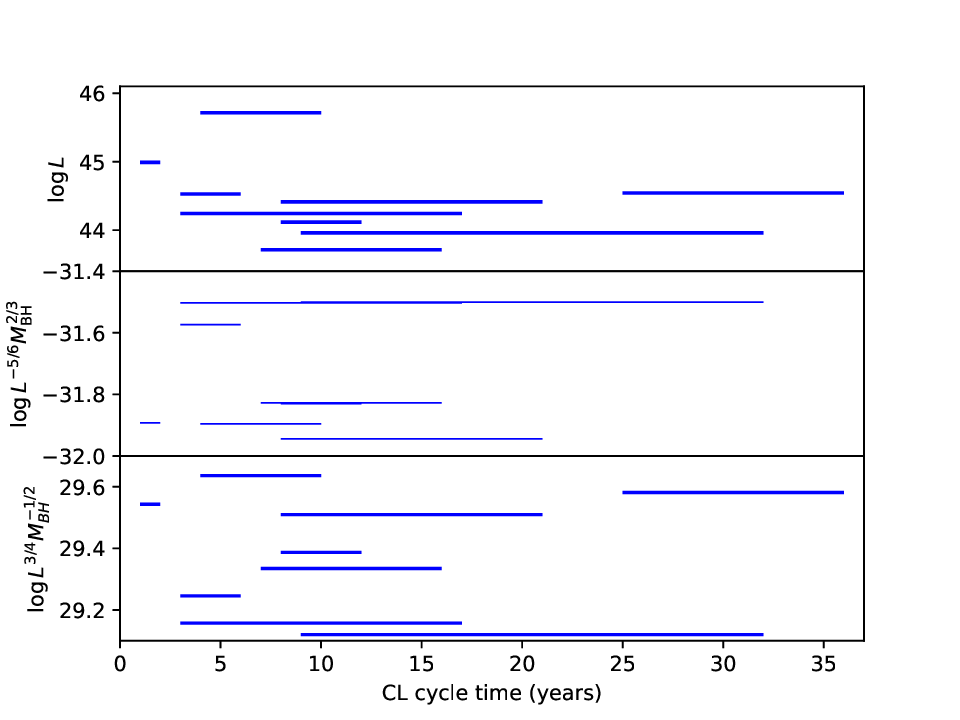}
\caption{The same as the upper panel in Figure 2, but for $t_{\mathrm{th}}\propto L^{1/2}$ 
(top panel), 
$t_{\mathrm{infl}}\propto L^{-5/6}M_{\mathrm{BH}}^{2/3}$ (middle panel) and 
$t_{\mathrm{cross}}\propto L^{3/2}M_{\mathrm{BH}}^{-1/2}$ (bottom panel), respectively.
\label{fig:general}}
\end{figure}

\subsection{Mass Transfer Instability}

Instead of the DI model with a constant mass transfer to an accretion
disk, the disk could become bright or faint due to a variation of the 
matter arriving at the disk. The latter has been originally proposed to explain the 
bursts or superbursts of dwarf novae, although it has been ruled out by a contradiction 
with observations (see Warner 1995 for a review).

In the classical Bondi accretion model (Bondi 1952), the characteristic time scale of accretion is found to
related with $M_{\mathrm{BH}}$ as (See Appendix D.1 for the details)
\begin{equation}
  \Delta t\approx3500\bigg(\frac{L_{\mathrm{bol}}}{L_{\mathrm{Edd}}}\bigg)^{3/2}
   \bigg(\frac{M_{\mathrm{BH}}}{10^7M_{\odot}}\bigg)^{-1/2}\ \mathrm{yr}
\end{equation}
where the sound speed and radius of the gas cloud surrounding the central SMBH
are adopted to be $10\mathrm{km\ s^{-1}}$ and to be $\propto L^{1/2}$, respectively. 
This relationship is over plotted in Figure 2 by a set of lines with different 
$\lambda=L_{\mathrm{bol}}/L_{\mathrm{Edd}}$. One can see from the figure 
the eleven RCL-AGNs well follow the relationships predicted from the Bondi 
accretion model with an $L_{\mathrm{bol}}/L_{\mathrm{Edd}}$ between 0.01 and 0.1.
In fact, by being biased against high $L_{\mathrm{bol}}/L_{\mathrm{Edd}}$, 
a majority of CL-AGNs identified so far are found to have $L_{\mathrm{bol}}/L_{\mathrm{Edd}}$ within
0.01 and 0.1 in their ``turn-on'' state (e.g., MacLeod et al. 2019; Wang et al. 2019, 
2023; Zeltyn et al. 2024). Based on the well-documented $M_{\mathrm{BH}}-M_{\mathrm{bugle}}$
relationship (e.g., Schutter et al. 2019), we propose that a comparable 
$L_{\mathrm{bol}}/L_{\mathrm{Edd}}$ 
can be resulted from the Bondi accretion model as long as the ratio of the total gas supply to 
the $M_{\mathrm{bugle}}$ is roughly a constant of $\sim10^{-11}$, which corresponds to a total 
gas mass of $\sim0.2M_\odot$ for a $10^{8}M_\odot$ SMBH (See Appendix D.2 for the details).

The proposed mass transfer scenario means that the CL phenomenon is resulted from a 
variation of accretion rate 
related to a sudden change of circumnuclear gas supplying or amount of the gas arriving at the 
disk, instead of just an instability occurring 
on a steady accretion disk. This sudden change of gas supplying coincides with the study recently reported in Wang et al. (2023), in which, because
of the associated intermediate-aged stellar populations (see also in Liu et al. 2021; 
Jin et al. 2022), CL phenomenon is proposed to
occur at a transition between the ``feast'' and ``famine'' SMBH fueling stages that 
are profiled by Kauffmann \& Heckman (2009). 
In addition, Dodd et al. (2021) points out that CL-AGNs tend to reside in 
high-density pseudobulges located in the so-called ``green valley'' (see also in 
Liu et al. 2021). The 
``green valley'' is believed to be a transition region between active star-forming 
galaxies and inactive elliptical galaxies. 
In the phase transition, 
because the gas reservoir is almost consumed by the previous steady accretion,
the SMBHs are instead fed by unsustainable and unstable gas supply resulted from 
slow stellar winds generated by evolved stars or by inward mass transport triggered 
by either minor mergers or multiple collisions of cold-gas clumps in the 
intergalactic medium (e.g., Kauffmann \& Heckman 2006; Davies et al. 2007; Gaspari et al. 2013; and see reviews in Heckman \& Best 2014).  
The scenario of gas supply from evolved circumnuclear stars is supported 
by larger metallicity in the BLR than NRL, which is recently revealed by reproducing the 
published observations by updated photoionization models in Huang et al. (2023).

Our proposed scenario does not entail the spontaneous development of an accretion disk surrounding the central SMBH solely through Bondi accretion. Instead, we suggest that the accretion rate of a SMBH could be boosted by material falling onto the disk, a process influenced not only by the corresponding physical mechanisms but also by the gas reservoir on a larger scale ($\sim$kpc). 

A similar concept has recently been put forward by Veronese et al. (2024) to explain the long-term variability of the CL AGN Mrk \,1018 across multiple wavelengths. The authors posit that the CL phenomenon arises from disk instability due to gas deposition resulting from the  fly-by clouds that sink onto the central region of the AGN through various mechanisms, including either cold chaotic accretion or mergers (refer to Ostrike 1999; Kormendy \& Ho 2013; Heckman \& Best 2014; Maccagni et al. 2021; Liu et al. 2021).

On a theoretical basis, although the gas with a small angular momentum falls almost freely 
to the central SMBH from the Bondi accretion radius, an accretion disk is argued to be formed within
the circularization radius $R_\mathrm{C}=R_\mathrm{B}(\Omega_\mathrm{B}/\Omega_\mathrm{K})^2$, where 
$\Omega_\mathrm{B}$ and $\Omega_\mathrm{K}$ ($\Omega_{\mathrm{B}}\ll\Omega_{\mathrm{K}}$) are the angular velocities
of the gas and the Keplerian velocity at the Bondi accretion radius, respectively. In addition,
Ram{\'\i}rez-Vel{\'a}squez et al. (2019) demonstrate a modification of the dynamics of the non-isothermal Bondi accretion flow due to an interaction between the falling material and the emission from the inner accretion disk. On the observational front, signatures of inflows have been unambiguously identified 
at different scales in 
a number of AGNs through the redshifted absorptions (e.g., Chen et al. 2022; Zhou et al. 2019; Shi et al. 2016; Rubin 2017 and references therein). Especially, fast inflows with velocities comparable to the freefall speed have been detected at the outer accretion disks of a few quasars (Zhou et al. 2019).



We argue that the Bondi accretion scenario is helpful for explaining the rare detection of 
CL phenomenon of narrow-line Seyfert 1 galaxies (NLS1s). Up to date, there are 
only five CL-NLS1s (e.g., Oknyansky et al. 2018; MacLeod et al. 2019; 
Frederick et al. 2019; Hon et al. 2022; Xu et al. 2023). 
Taking into account of their small SMBHs ($M_{\mathrm{BH}}\sim10^7M_\odot$) and high accretion 
rate close to the Eddington limit (e.g., Boroson 2002; Grupe 2004; Xu et al. 2012; 
see Komossa 2008 for a review), the rarity of CL-NLS1s could be explained in terms of the 
proposed Bondi accretion scenario because of the very long CL cycle time ($\sim10^3$yr) predicted
by Eq. (3). 

Finally, further observations with high spatial resolution by such as VLBI experiments and simulation of dynamics within $\sim1$ to 100pc will be helpful 
for validating the proposed scenario.

\rm


\section{Conclusions}

In our investigation into the origin of the CL phenomenon in certain SMBHs, we have specifically focused on RCL-AGNs. Our findings indicate that larger SMBHs tend to have shorter CL cycle times, while smaller SMBHs have longer CL cycle times. This trend is challenging to explain
CL phenomenon in terms of the well-known AGN's DI model. By comparing observations with theoretical predictions, we propose that the CL phenomenon is likely a result of a variation in accretion rate
resulted from a sudden change of the supply of circumnuclear gas. Identifying new RCL-AGNs and expanding the RCL-AGN sample will allow us to validate the observed trend and the proposed scenario, and further enhance our understanding of the mechanisms behind the CL phenomenon.

\acknowledgments

We thank the anonymous referee for the helpful comments that improved our 
study significantly.
This study is supported by the National Natural Science Foundation of China (under grants 12173009, 12273054 and 12233007) and the Natural Science Foundation of Guangxi (2020GXNSFDA238018). 
We would like to thank S. Komossa and E. L. Qiao for very useful suggestions and comments.
We acknowledge the support of the staff of the Xinglong 2.16m telescope. This work was partially supported by the Open Project Program of the 
Key Laboratory of Optical Astronomy, National Astronomical Observatories, Chinese Academy of Sciences. 
This study used the NASA/IPAC Extragalactic Database (NED), which is operated by the Jet Propulsion
Laboratory, California Institute of Technology.

\vspace{5mm}
\facilities{NAOC Xinglong 2.16 m telescope}
\software{IRAF (Tody 1986, 1992), MATPLOTLIB (Hunter 2007)}
%
\clearpage

\appendix
\section{Spectroscopic Observations and Data Reduction for RCL-AGNs} 
New long-slit spectra of five RCL-AGNs were obtained by the Beijing Faint Object Spectrograph and Camera (BFOSC) mounted 
on the 2.16~m telescope (Fan et al. 2016) at the Xinglong Observatory of the National Astronomical Observatories, Chinese Academy
of Sciences (NAOC) in several runs from 2019 to 2023.  
The G4 grism and a long slit of width 1.8\arcsec\
oriented in the north–south direction were used in the observations, which leads to a spectral 
resolution of $\sim10$\AA\ and a wavelength coverage of 3850—8200\AA.
Wavelength calibration was carried out with spectra of iron-argon comparison lamps. In order to minimize the effects of atmospheric 
dispersion, all spectra were obtained as close to the meridian as possible. 
In most cases, each object was observed successively twice or more in each night. The
multi-exposures were combined prior to extraction to enhance
the S/N ratio and to eliminate the contamination
of cosmic rays easily.
The total exposure time of individual object ranges from 1200 to 4800~s, 
depending on the brightness and weather condition. 
The log of the observations are presented in Table 2. 

 \begin{table*}
       \centering
         \caption{Observation log of the five RCL-AGNs}
         \footnotesize
         \label{tab:example_table}
         \begin{tabular}{cccccc} 
         \hline
         \hline
         Object & R.A. &  DEC & $z$ &  Date & Exposure time\\
               &  degree   &  degree   &     &  & seconds \\
           (1) & (2) & (3) & (4) & (5) & (6) \\     
         \hline     
        Mrk\,590 & 33.639844 & -0.766693 & 0.02638 & 2019-11-28 & $2\times1200$\\
                 &           &           &         & 2020-01-22 & $2\times1600$\\
                 &           &           &         & 2024-01-09 & $2000$\\
        Mrk\,1018 & 31.566625 & -0.291444 & 0.04296 & 2019-11-26 & $2\times1800$\\
                 &           &           &         & 2021-01-12 & 1800\\
                 &           &           &         & 2021-01-12 & 1200\\
        UGC\,3223 & 74.789126 & 4.975002 & 0.01562 & 2020-01-01 & $2\times1800$\\
                 &           &           &         & 2020-01-21 & $2\times1800$\\
                 &           &           &         & 2021-02-04 & $2\times1500$\\
                 &           &           &         & 2022-02-22 & $2\times1800$\\
                 &           &           &         & 2023-02-20 & $2\times1200$\\
        NGC\,4151 &  182.635745 & 39.405730 & 0.003326& 2024-01-09 & 480 \\ 
        3C\,390.3 & 280.537458 & 79.771424 & 0.05613 & 2023-05-17 & 2000\\
        NGC\,7603 & 349.735906 & 0.243952 & 0.02876 & 2019-11-26 & $2\times2400$\\
                 &           &           &         & 2020-10-15 & $3\times1200$\\
                 &           &           &         & 2023-09-22 &  1200\\
         \hline
         \end{tabular}
         \tablecomments{Columns (1): Object IAU name; Columns (2) and (3): equatorial  coordinates in unit of degree at epoch of J2000; Column (4): Redshift; 
         Columns (5): Observation date in UTC; Column (6): Exposure time in unit of second.}
 \end{table*}

We reduced the raw images and extracted 1D spectra by utilizing IRAF\footnote{ IRAF is distributed by NOAO, which is operated by AURA, Inc., under cooperative 
agreement with the U.S. National Science Foundation (NSF).} packages (Tody 1986, 1992) and standard procedures for bias subtraction and flat-field correction.

All of the extracted 1D spectra were then calibrated in wavelength and in flux with the spectra of the comparison lamps and the Kitt Peak National Observatory
standard stars (Massey et al. 1988). 
The accuracy of the wavelength calibration is better than 2\AA.\ The standard-star spectra  were used to remove the telluric A-band (7600—7630\AA) and
B-band (around 6860\AA) absorption caused by atmospheric $\mathrm{O_2}$ molecules. 
Each calibrated spectrum was then corrected for Galactic extinction according to the color excess $E(B-V)$ taken from the Schlafly \& Finkbeiner (2011) Galactic 
reddening map, in which the $R_V = 3.1$ extinction law of our Galaxy (Cardelli et al. 1989) is adopted.
All the spectra were then transformed to the rest frame according to their redshifts.
The rest-frame spectra are displayed in Figure 4.

\begin{figure}[h]
\includegraphics[width=1.0\textwidth]{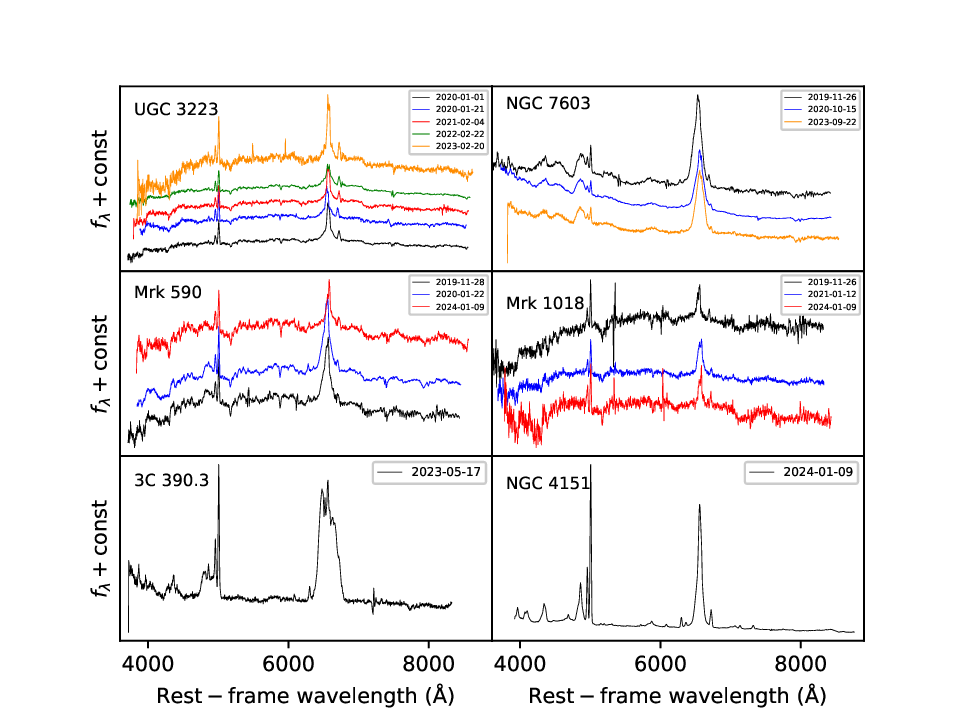}
\caption{Spectral sequence of the five RCL-AGNs obtained by the NAOC Xinglong 2.16m telescope from 2019 to 2023. In each panel, the spectra in rest-frame are vertically shifted by an arbitrary 
amount for clarity.
\label{fig:general}}
\end{figure}
 
\section{Characteristic Time Scales of Radial Viscous Inflow and Thermal Instability}

In the SSD model, by assuming the local emission is thermal with different temperature 
and all the heat is radiated, the effective temperature of the disk photosphere is 
\begin{equation}
  T_{\mathrm{eff}}(r)\propto \eta^{-1/4}\bigg(\frac{L}{L_{\mathrm{Edd}}}\bigg)^{1/2}L^{-1/4}x^{-3/4}
\end{equation}
after ignoring the general relativity effect, where $\eta\sim0.1$ is the accretion efficiency, $x=r/r_{\mathrm{g}}$ the dimensionless distance 
from the central SMBH and $r_{\mathrm{g}}=GM_{\mathrm{BH}}/c^2$ the 
Schwarzchild radius. 

Based on the Wien's displacement law $\lambda_{\mathrm{max}}T=$constant of a blackbody, 
the dependence of $T_{\mathrm{eff}}\propto x^{-3/4}$ means the light curve observed at
different wavelengths are mainly contributed by disk rings with different radius.   
Re-arranging EQ. (B1) leads to 
\begin{equation}
  x\propto T_{\mathrm{eff}}^{-4/3}L^{1/3}M_{\mathrm{BH}}^{-2/3}
\end{equation}
or 
\begin{equation}
  r\propto T_{\mathrm{eff}}^{-4/3}L^{1/3}M_{\mathrm{BH}}^{1/3}
\end{equation}
which can be simplified to be 
\begin{equation}
  r\propto L^{1/3}M_{\mathrm{BH}}^{1/3}
\end{equation}
for a fixed $T_{\mathrm{eff}}$.

The characteristic time scale is given by 
\begin{equation}
    t_{\mathrm{infl}} = 6.5\bigg(\frac{\alpha}{0.1}\bigg)^{-1}\bigg(\frac{L/L_{\mathrm{Edd}}}{0.1}\bigg)^{-2}\bigg(\frac{\eta}{0.1}\bigg)^2\bigg(\frac{r}{r_{\mathrm{g}}}\bigg)^{7/2}\bigg(\frac{M_{\mathrm{BH}}}{10^8~{\rm M}_\odot}\bigg)~\mathrm{yr}\,
\end{equation}
for the radial viscous inflow scenario (LaMassa et al. 2015)
and
\begin{equation}
\small
 t_{\mathrm{th}} \approx \frac{1}{\alpha\Omega_{\mathrm{K}}}\\
 =2.7\bigg(\frac{\alpha}{0.1}\bigg)^{-1}\bigg(\frac{r}{10^{16}\,\mathrm{cm}}\bigg)^{3/2}\bigg(\frac{M_{\mathrm{BH}}}{10^8~{\rm M}_\odot}\bigg)^{-1/2} \mathrm{yr}, 
\end{equation}
for the thermal instability scenario (Siemiginowska et al. 1996), 
where $\alpha\sim0.1$ is the “viscosity” parameter.

Substituting Eq. B4 into the above two equations finally leads to 
\begin{eqnarray}
    t_{\mathrm{infl}} & \propto & L^{-5/6}M_{\mathrm{BH}}^{2/3}= L^{-1/6}\bigg(\frac{L}{L_{\mathrm{Edd}}}\bigg)^{-2/3} \\
    t_{\mathrm{th}} & \propto & L^{1/2} 
\end{eqnarray}

\section{Characteristic Time Scales of Obscuration}

For the obscuration material orbiting outside the broad-line region (BLR) on a Keplerian orbit,
the crossing time can be estimated from (LaMassa et al. 2015)
\begin{equation} 
  t_{\mathrm{cross}}=0.07\bigg(\frac{r_{\mathrm{orb}}}{\mathrm{1ld}}\bigg)^{3/2}\sin^{-1}\bigg(\frac{r_{\mathrm{src}}}{r_{\mathrm{orb}}}\bigg)
  \bigg(\frac{M_{\mathrm{BH}}}{10^8M_\odot}\bigg)^{-1/2}
\end{equation}
where $r_{\mathrm{orb}}$ and $r_{\mathrm{src}}$ are the orbital radius of the obscuration
material and the size of BLR, respectively.

With the classical $R-L$ relation $r_{\mathrm{src}}\propto L^{1/2}$ (e.g., Kaspi et al. 2000, 2005; Peterson et al. 2004; 
Bentz et al. 2013; Du et al. 2014, 2015; Du \& Wang 2019), we have 
\begin{equation}
   t_{\mathrm{cross}}\propto L^{3/4}M_{\mathrm{BH}}^{-1/2}
\end{equation} 
after adopting $r_{\mathrm{orb}}=r_{\mathrm{src}}$.

\section{Bondi Accretion}
\subsection{Characteristic Time Scale}
By consuming gas with a total mass of $\Delta M$ at a rate of $\dot{M}$, 
the characteristic accretion time scale can be written as
\begin{equation}
  \Delta t=\frac{\Delta M}{\dot{M}}
\end{equation} 
In the Bondi accretion model, the accretion rate of a point source (with a mass 
of $M_{\mathrm{BH}}$) embedded in a uniform gas cloud is (Bondi 1952) 
\begin{equation}
  \dot{M}_{\mathrm{bondi}}=\frac{4\pi G^2M_{\mathrm{BH}}^2\rho}{c_{\mathrm{s}}^3}
\end{equation}
where $G$, $\rho$ and $c_{\mathrm{s}}$ are the gravity constant, gas density and sound speed,
respectively. 
With $M=\rho V$, where $V=\frac{4\pi}{3}r^3$ is the volume of gas and $r$ the radius of the 
gas cloud surrounding the central SMBH (or the Bondi accretion radius), we find that
\begin{equation}
  \Delta t=\frac{\Delta M}{\dot{M}_{\mathrm{bondi}}}=\frac{\rho V}{\dot{M}_{\mathrm{bondi}}}=
  \frac{c_{\mathrm{s}}^3V}{4\pi G^2M_{\mathrm{BH}}^2}=
  \frac{c_{\mathrm{s}}^3r^3}{3G^2M_{\mathrm{BH}}^2}
\end{equation}
or numerically  
\begin{equation}
  \Delta t=36\bigg(\frac{c_{\mathrm{s}}}{\mathrm{10\ km\ s^{-1}}}\bigg)^3
  \bigg(\frac{r}{\mathrm{1pc}}\bigg)^3\bigg(\frac{M_{\mathrm{BH}}}{10^7\ M_\odot}\bigg)^{-2}
  \mathrm{yr}
\end{equation}

In our scenario, the accretion time scale $\Delta t$ is predicted to be sensitive to the 
accretion radius that is, however, hard to be determined accurately in the current stage.  
Several available options are discussed as follows.
The simplest way to determine the accretion radius is the
Bondi accretion radius that can be written as 
(Frank et al. 2002)
\begin{equation}
r_{\mathrm{acc}}=\frac{2GM}{c_{\mathrm{s}}^2}\simeq3\times10^{21}\bigg(\frac{M_{\mathrm{BH}}}{10^7M_\odot}\bigg)\bigg(\frac{T}{10^4\mathrm{K}}\bigg)^{-1}\ \mathrm{cm}
\end{equation}
for a given SMBH, where $c_{\mathrm{s}}\approx 10(T/10^4\ \mathrm{K})^{1/2}\ \mathrm{km\ s^{-1}}$ for both isothermal and adiabatic processes. \rm 
For the current sample with $M_{\mathrm{BH}}=10^{7-9}M_\odot$, the value of $r_{\mathrm{acc}}$ is predicted to be $\sim1-100$kpc by adopting $T=10^4$K. This value is, however, far beyond the 
size of AGN's narrow-line region (NLR, $\sim10^2$pc) whose gas kinematics is dominated by the 
host bugle.

Substituting the above equation into Eq. (D14) yields
\begin{equation}
   \Delta t=3.3\times10^{10}\bigg(\frac{T}{\mathrm{10^4\ \mathrm{K}}}\bigg)^{-3/2}
   \bigg(\frac{M_{\mathrm{BH}}}{10^7\ M_\odot}\bigg)
  \mathrm{yr}
\end{equation}
which allows us to argue against the use of the Bondi accretion radius for 
two reasons. At first,
the gas temperature being as high as $\sim10^{10}$K is required to reproduce the observed CL cycle timescale of years to decades. Secondly, $\Delta t$ is predicted to be proportional to 
$M_{\mathrm{BH}}$ by adopting a universal gas temperature, which is however inconsistent with our observations.

According to Eq. (D14),
the CL timescale $\Delta t$ is predicted to be $\sim$weeks ($\sim100$yr)
for the current RCL-AGN sample, when $r=r_{\mathrm{sub}}\sim0.1$pc ($r=r_{\mathrm{NLR}}\sim100$pc) is adopted. $r_{\mathrm{sub}}$ and $r_{\mathrm{NLR}}$ are the inner radius of the torus determined by dust sublimation and the typical radius of the NLR, respectively. Both radius have been found to relate with AGN's
luminosity as $r\propto L^{0.5}$ (e.g., Barvainis 1987; Laor \& Draine 1993; 
Baskin \& Laor 2018; Kishimoto et al. 2007; Mor et al. 2009).

In order to reproduce CL cycle time scale comparable to our observations, 
we alternatively assume the volume of the surrounding gas accreted by a SMBH 
is between that of torus (or BLR) and NLR, i.e., $r_{\mathrm{sub}}<r<r_{\mathrm{NLR}}$. 
Taking into account the fact that $r_{\mathrm{NLR}}$ differs from
$r_{\mathrm{sub}}$ by at least three orders of magnitude,
a logarithmic average, i.e., $r=\sqrt{r_{\mathrm{sub}}r_{\mathrm{NLR}}}\sim1-10$pc\footnote{Although this distance is out of the  gravitational potential of a SMBH, we believe it is still reasonable because of the well 
documented $M_{\mathrm{BH}}-M_{\mathrm{bugle}}$ relationship 
(e.g., Reines \& Volonteri 2015; Schutte et al. 2019).}, is used by us, which leads to 
\begin{equation}
   \Delta t \approx 3500\bigg(\frac{L_{\mathrm{bol}}}{L_{\mathrm{Edd}}}\bigg)^{3/2}
   \bigg(\frac{M_{\mathrm{BH}}}{10^7M_{\odot}}\bigg)^{-1/2}\ \mathrm{yr}
\end{equation}
\rm after adopting $r_{\mathrm{sub}}=0.47L_{46}^{1/2}$ pc (Kishimoto et al. 2007) and $r_{\mathrm{NLR}}=295L_{46}^{0.47\pm0.13}\simeq295L_{46}^{1/2}$ pc
(Mor et al. 2009), where $L_{46}=L_{\mathrm{bol}}/10^{46}\ \mathrm{erg\ s^{-1}}$.

By adopting a constant $L/L_{\mathrm{Edd}}=0.03$
(the average of current sample, see Table 1), we have a relationship:
\begin{equation}
  \log\bigg(\frac{M_{\mathrm{BH}}}{1M_\odot}\bigg)=9.5-2\log\bigg(\frac{\Delta t}{\mathrm{1 yr}}\bigg) 
\end{equation}
which is over-plotted in Figure 2 by the dot-dashed line.

\subsection{Luminosity}

The accretion luminosity of the Bondi accretion is 
\begin{equation}
  L_{\mathrm{acc}}=\eta\dot Mc^2=\eta c^2\frac{\Delta M}{\Delta t}
\end{equation}
where $\eta\sim0.1$ is the radiative efficiency in the rest-mass units. 

By supposing the total gas 
supply $\Delta M$ is related to the bulge mass $M_{\mathrm{bulge}}$ as $\Delta M=f_{\mathrm{gas}}M_{\mathrm{bulge}}$, we have 
\begin{equation}
 \Delta M\sim10^5f_{\mathrm{gas}}M_{\mathrm{BH}}
\end{equation}
after taking into account of the well known $M_{\mathrm{BH}}-M_{\mathrm{bugle}}$ relationship: 
$\log(M_{\mathrm{BH}}/M_\odot)=(1.24\pm0.08)\log(M_{\mathrm{bulge}}/10^{11}M_\odot)+(8.80\pm0.09)$
(Schutte et al. 2019).

Combining Eq. (D18) and Eq. (D19), the Eddington ratio can be written as 
\begin{equation}
  \frac{L_{\mathrm{bol}}}{L_{\mathrm{Edd}}}=10^{2}f_{\mathrm{gas}}\frac{\eta c^2}{\Delta t}
  =3\times10^9f_{\mathrm{gas}}\bigg(\frac{\eta}{0.1}\bigg)\bigg(\frac{\Delta t}{10\mathrm{yr}}\bigg)^{-1}
\end{equation}
Adopting the fiducal values of $\eta=0.1$ and $\Delta t=10$yr, a value of $L_{\mathrm{bol}}/L_{\mathrm{Edd}}=0.03$ can be found as long as a universal gas fraction of $f_{\mathrm{gas}}=10^{-11}$.


\end{document}